# Quantum criticality in ferroelectrics

S.E. Rowley[1], L.J. Spalek[1], R.P. Smith[1], M.P.M. Dean[1], G.G. Lonzarich[1], J.F. Scott[2] and S.S. Saxena[1]

1. Cavendish Laboratory, University of Cambridge, J.J. Thomson Avenue, Cambridge, CB3 0HE, UK
2. Department of Earth Sciences, University of Cambridge, Downing Street, Cambridge, CB2 3EQ, UK

**Materials tuned to the neighbourhood of a zero temperature phase transition often show the emergence of novel quantum phenomena. Much of the effort to study these new effects, like the breakdown of the conventional Fermi-liquid theory of metals has been focused in narrow band electronic systems. Ferroelectric crystals provide a very different type of quantum criticality that arises purely from the crystalline lattice. In many cases the ferroelectric phase can be tuned to absolute zero using hydrostatic pressure or chemical or isotopic substitution. Close to such a zero temperature phase transition, the dielectric constant and other quantities change into radically unconventional forms due to the quantum fluctuations of the electrical polarization. The simplest ferroelectrics may form a text-book paradigm of quantum criticality in the solid-state as the difficulties found in metals due to a high density of gapless excitations on the Fermi surface are avoided. We present low temperature high precision data demonstrating these effects in pure single crystals of $SrTiO_3$ and $KTaO_3$. We outline a model for describing the physics of ferroelectrics close to quantum criticality and highlight the expected $1/T^2$ dependence of the dielectric constant measured over a wide temperature range at low temperatures. In the neighbourhood of the quantum critical point we report the emergence of a small frequency independent peak in the dielectric constant at approximately 2 K in $SrTiO_3$ and 3 K in $KTaO_3$ believed to arise from coupling to acoustic phonons. Looking ahead, we suggest that ferroelectrics could be used as systems in which to controllably build in extra complexity around the quantum critical point. For example, in ferroelectric or anti-ferroelectric materials supporting mobile charge carriers, quantum paraelectric fluctuations may mediate new effective electron-electron interactions giving rise to a number of possible states such as superconductivity.**

The study of quantum matter at low temperatures has given rise to a fascinating and often surprising catalogue of phenomena important to our understanding of nature[1] and to technological development[2]. In particular, the study of materials close to a continuous low temperature phase transition or so called quantum critical point forms an important branch of research within condensed matter physics. A chief reason for this is that close to such a transition, materials become highly degenerate and new states of matter are frequently found to emerge. In fact, it turns out that many materials end up being close to or within the quantum critical regime. This is because quantum critical phenomena can affect materials over a wide range of temperatures, pressures and other variables. In electrically conducting materials, the standard model of the metallic state, Landau's Fermi liquid theory is seen to breakdown close to the low temperature boundary between a magnetic and paramagnetic phase and is replaced with other forms of novel quantum liquid. For example, in some weakly magnetic cubic d-metals evidence is growing for the so-called marginal Fermi-liquid, where quantities such as the resistivity, susceptibility and heat capacity take on particularly unusual forms[3]. As well as these exotic metals, completely new broken symmetry states have been observed such as inhomogeneous magnetic phases[4,3] and unconventional superconductors where the electrons are thought to be paired via magnetic fluctuations[5,6,7]. However in metals, the applicability of simple models for the quantum critical point are complicated by a variety of factors, a major one being that the metallic vacuum is characterized by a Fermi surface around which exists a multitude of gapless excitations. On the other hand, insulating ferroelectrics are in some respects simpler. The low temperature properties close to a ferroelectric phase transition can be described by considerations of bosonic modes alone, that is quantized waves of lattice vibrations or phonons. The quantum critical effects arise from the lattice system and in the simplest cases there is no damping of the phonons due to mobile electrons. In this paper we present high-precision experimental data as well as a simple model in order to understand the low temperature properties of incipient ferroelectrics. We outline connections to earlier work and other fields and suggest that the simplest ferroelectric materials present a text-book paradigm of quantum criticality in the solid-state. With this in mind we also point out that there are interesting extensions to the simplest cases that can come into play. For example, subtleties involving long-range interactions associated with the electrical dipole moments can be important in some materials. In others, interactions of the critical phonon modes with other phonon branches or coupling of the ferroelectric properties to itinerant electrons and other degrees of freedom can lead to a rich phase diagram close to the quantum critical point. It will be seen that fluctuations in the polarization of cubic ferroelectric materials close to quantum criticality typically exist in four dimensions. The usual Curie-Weiss form of the electrical susceptibility $\chi^{-1} \sim T$ changes into a radically different form, $\chi^{-1} \sim T^2$ in the quantum critical regime. Other quantities such as the thermal expansion coefficient and soft-mode frequencies are also expected to vary differently from those observed in the vast majority of ferroelectric crystals, indicating that the solid-state lattice is vibrating in an unconventional manner. Since fluctuations of the ferroelectric order parameter are typically occurring at the marginal dimension ($d_{eff}$ = 4), logarithmic corrections to mean-field theory should be included in close proximity to the quantum critical point. Seminal experimental work was carried out observing the predictions of renormalization group calculations in magnetic systems near to classical phase transitions[8]. Close enough to a ferroelectric quantum critical point there is an opportunity to measure the related quantum renormalization group corrections that are not normally observed in other quantum critical systems due to their respective fluctuations existing in dimensions greater than marginal. After outlining a simple model to understand the properties of ferroelectrics close to quantum criticality and presenting the temperature-pressure phase diagram one might expect, we show experimental results in the perovskites $SrTiO_3$ and $KTaO_3$. We find a striking agreement between the predictions of the model and susceptibility measurements at ambient pressure and observe the effect of quantum critical fluctuations as defined later up to approximately 50 K. Of particular interest is the appearance of a frequency independent peak in the dielectric susceptibility at approximately 2 K in $SrTiO_3$ and 3 K in $KTaO_3$ in the neighbourhood of the quantum critical point. The origin of this new low temperature regime in unbiased and unstrained pure single crystals may arise from coupling of the quantum fluctuations with sound waves (acoustic phonons) and warrants detailed investigation in future work.

**The Border of Ferroelectricity**

The microscopic origin of ferroelectricity in the kind of systems we wish to consider is related to an instability in the lattice of the high temperature, high symmetry phase. The forces that lead to the instability are anharmonic and lead to the softening of an optical phonon mode which, depending on the particular crystalline details, can produce ferroelectric or non-ferroelectric structural phase transitions[9,10,11]. In this paper we consider the class of displacive ferroelectrics where the lowest transverse optical phonon mode becomes soft at the zone centre, eventually freezing out with zero frequency at a critical temperature, $T_C$. Below this temperature, the system undergoes breaking of lattice inversion symmetry and a spontaneous polarization starts to grow. It is interesting to note that in calculations involving periodic boundary conditions without the presence of a surface, the traditional definition of polarization as being related to the dipole moment of the unit cell is not always appropriate. This is particularly evident when the dipoles are not generated purely from charges at particular ionic sites but are connected to the electronic distribution spread out within a unit cell. A number of papers have pointed out that the definition of polarization in these conditions is most conveniently expressed as a charge current via the quantum mechanical phase of the wavefunctions[12,13]. Thus in this sense ferroelectricity is inherently a quantum effect, in that the order parameter is expressed as a geometric (Berry) phase of the underlying wavefunctions. The quantum effects discussed in this paper are of a very different kind and come into play close to a quantum critical point where the temperature scale of the system is much less than the Debye temperature of the optical phonon branch. We have in mind materials that are driven to the border of low temperature ferroelectricity via the use of hydrostatic pressure or some other control parameter such as chemical or isotopic substitution. In some cases, pure materials fall naturally in this region at ambient pressure as exemplified by $SrTiO_3$ and $KTaO_3$ and which we investigate here. In the literature materials of this kind are widely referred to as 'quantum paraelectrics' after the work of Müller et al[14].

Earlier studies on the theory of quantum criticality in ferroelectrics can be found in the papers of Rechester[15], Khmelnitskii and Shneerson[16,17] and references therein. In these papers they use diagrammatic methods to understand the physics around the quantum critical point including the logarithmic corrections associated with marginal dimensionality. Later supporting work was carried out by Schneider, Beck and Stoll[18] as well as more recently by Roussev and Millis[19] who address the problem using modern renormalization group techniques. Motivated by our work, Pálová et al[20] have also reviewed the theory of quantum paraelectrics in some detail.

The model we sketch here is a self-consistent phonon (SCP) treatment of the soft transverse-optic phonon branch that leads to ferroelectricity and is analogous to spin fluctuation theories used for weak itinerant magnets such as those by Lonzarich[21,22] and Moriya[23]. Using the model and our measurements we are able for the first time to make careful quantitative comparisons between experiment and theory with no adjustable parameters. The calculations we employ involve both the temporal and spatial fluctuations of the order parameter that are key in understanding the quantum critical behaviour and give rise to the unusual temperature dependencies of measured quantities. Quantum criticality theory permits the addition of (imaginary) time as an extra dimension to the usual three spatial dimensions. At zero temperature this extra dimension is infinitely ranged, and is treated on an equal footing with spatial dimensions in the paradigm of the general theory of critical phenomena. The extra dimension is also important over a range of pressures and crucially at finite temperatures but is now of finite size[24]. The particular temperature range over which quantum critical fluctuations are important depend on the particular system or compound in question. The model below allows one to predict quantitatively the temperature range over which, for example, the dielectric constant will be modified by quantum critical fluctuations. It depends on four material specific, non-adjustable, zero-temperature parameters that can be determined either from independent experiments or via first principles calculations. We find from theory and measurement that quantum critical fluctuations in the sense defined below persist in the materials investigated here up to the relatively high temperature of 50 K, above which we observe a cross-over to classical behaviour.

We consider the simplest case of a low temperature paraelectric material with cubic crystal symmetry close to a second-order or weakly first-order displacive transition. The order parameter in the model is the coarse grained polarization $P$ defined as the average dipole moment density over a mesoscopic volume. The key point regarding ferroelectrics is that as well as short ranged lattice interactions, there exist long range dipole-dipole forces. The displacement of ions about equilibrium is small compared to the unit cell size and for materials with no preferred direction of the polarization, the order parameter has a continuous (Heisenberg like) symmetry. The fluctuations in the polarization can be split into one longitudinal and two transverse components with respect to the direction of wave propagation. The effect of the dipole forces is to stiffen or gap out the longitudinal components keeping the remaining transverse ones free. These transverse parts alone become soft at criticality and dominate the physics and as such after neglecting the longitudinal part, the dipole interactions can be ignored. In some respects a more idealised case would be the non-polar displacive quantum critical point where the complications arising from dipole interactions are avoided. It turns out however that the interesting cases are the ones that have dipoles since it is in these materials where the susceptibility can be measured with very high precision and are of importance for applications. The effects of dipoles only lead to quantitative rather than qualitative changes in the limits of interest. As a side point we note that in cases when there is an easy axis for the polarization direction such as in uniaxial or strongly tetragonal materials, dipole interactions considerably change the physics. It can be shown that the dimensionality is effectively raised by one in both quantum and classical regimes and leads to different temperature dependences of the dielectric constant to the results reported here. All of these cases including a detailed examination of dipole interactions have been discussed elsewhere and can be found for example in references[25,9,17,19]. The expected logarithmic corrections for $d_{\text{eff}} = 4$ have been measured at the classical uniaxial ferroelectric critical points in trisarcosine calcium chloride[26] and $RbH_2PO_4$ [27]. Also for the uniaxial (dipolar) magnetic classical critical point in $LiTbF_4$ [8].

With these points in mind we consider for the Heisenberg case the zero temperature equation of state for a cubic crystal that relates the space and time varying electric field $E$ to the space and time varying polarization $P$. We keep terms up to forth order in the free energy and for simplicity assume isotropic interactions in an isotropic medium such that the static zero-temperature equation of state in *SI* units is $\varepsilon_0 E[P] = aP + bP^3 - c\nabla^2 P$ plus dipole interactions, which having eliminated the longitudinal component we neglect from now on. We have assumed that after dropping the long-range interactions a power series expansion in $P$ of the equation of state is valid. The parameters $a$, $b$ and $c$ are zero temperature material specific constants. The gradient term is related to the dispersion of the soft modes and the size of $c$ controls the stiffness of the system to spatial modulations in the polarization.

The dynamics of the fluctuations key in understanding the quantum critical behaviour are determined by the wavevector $q$ and frequency $\omega$ dependent dielectric susceptibility defined through the Fourier components of $E$ and $P$, $\chi_{q\omega} = \varepsilon_0 P_{q\omega}/E_{q\omega}$ and since the phonon modes are propagating this takes the form of a harmonic oscillator $\chi_{q\omega}^{-1} = \chi_q^{-1}(1 - \omega^2/\Omega_q^2)$ with negligible damping at small wavevectors and frequencies. The transverse-optic soft-mode dispersion is given by $\Omega_q^2 = \Delta^2 + v^2 q^2 = (\Delta^2/a)\chi_q^{-1}$ which is physically applicable for the long wavelength or small $q$ ($=|q|$) behaviour that dominates the physics close to the critical point and $\chi_q^{-1} = \chi^{-1} + cq^2$. $\chi^{-1}$ is the zero frequency, zero wavevector inverse susceptibility and is related to the measured dielectric constant $\varepsilon$ by $\chi^{-1} = (\varepsilon - 1)^{-1}$. Note that the gap $\Delta$ in the optical mode dispersion falls to zero at criticality resulting in linearly dispersing phonons with speed $v$, $\Omega_q(T = T_C) = vq$. The zero temperature values of $\Delta$ and $v$ can be determined for example by inelastic neutron scattering experiments and are used as fixed parameters in this theory. At the quantum critical point $\Delta = 0$, but at zero temperature away from the critical point $\Delta$ is finite. The power of $q$ in the dispersion relation at quantum criticality gives $z$, the *dynamical exponent*, and therefore in our case $z = 1$. This leads to fluctuations of the order parameter in an effective four dimensional space, $d_{eff} = d + z = 4$ for three dimensional materials ($d = 3$). This can be contrasted with itinerant ferromagnets and antiferromagnets where $d_{eff} = 6$ and $d_{eff} = 5$ respectively[23]. As mentioned earlier, for uniaxial ferroelectrics there is a further increase in dimensionality $d_{eff} = d + 1 + z$ which we do not consider further here.

Anharmonic interactions between the different wavevector modes of the transverse-optical branch we are treating leads to a renormalization of the coefficient of the linear term in the equation of state $a \rightarrow A$ and in the mean-field approximation this is $A = a + Nb\overline{p^2(T)}$ where $p$ is the stochastic fluctuating polarization about the mean for one of the equivalent transverse mode components. The number $N$ is chosen to take into account the vector nature of the model and the fact that only transverse optic phonons become soft such that in this case $N = 10/3$. Apart from possible logarithmic corrections, the mean-field approximation is valid for $d_{\text{eff}} \geq 4$, that is, equal to and above the upper critical dimension. Note that in the paraelectric state $A$ is simply the inverse susceptibility $\chi^{-1} = A$ and gains its temperature dependence through the variance of the polarization, $p$ alone. $A$ is positive at zero temperature for a paraelectric and negative for a ferroelectric ground state. The coefficient of the cubic term in the equation of state $b$ which is positive for a material in the vicinity of a second-order phase transition is assumed to be temperature independent. In an extension of the model to higher powers in $P$ in the equation of state, $b$ itself would also be renormalized and gain its own temperature dependence. This is especially necessary in cases when $b$ is very small or close to a first-order transition whereby $b$ is negative. The variance of $P$ is given by the fluctuation-dissipation theorem[28], $\overline{p^2(T)} = (\varepsilon_0/V)\sum_{q<q_c}(2\hbar/\pi)\int_0^\infty (n_\omega + 1/2)Im(\chi_{q\omega})d\omega$, where $V$ is the volume. The thermal properties of $\overline{p^2(T)}$ come from the Bose function $n_\omega = (e^{\hbar\omega/k_B T} - 1)^{-1}$. We say that there are quantum critical fluctuations in $\overline{p^2}$ when close to a quantum critical point and the temperature is much less than the Debye temperature of the optical phonon mode, $\theta_{D,TO}$. This is when the high temperature approximation of the Bose function cannot be made. Substituting for the imaginary part of $\chi_{q\omega}$ in the limit of negligible damping and integrating over frequency, leads to the following expression for the inverse susceptibility, $\chi^{-1} = a + ((10/3)b\Delta^2\varepsilon_0\hbar/aV)\sum_{q<q_c}(n_{\Omega_q} + 1/2)/\Omega_q$. The summation over $q$ is terminated at a cut-off wavevector $q_c$ that is set by the coarse graining length assumed in the model. This equation includes the dynamics of the system through the frequency dependent terms as well as quantum zero-point fluctuations. It depends on four zero-temperature non-adjustable parameters $a$, $b$, $\Delta$ and $v$ ($c = av^2/\Delta^2$) that can be found from experiment or first principles calculations. The results are only very weakly dependent on the cut-off $q_c$ and in the numerical calculations we take this to be the Brillouin zone boundary. The equation is self-consistent as the susceptibility appears on both sides and in general must be solved numerically. Note that the measured value of *a* includes the zero-temperature zero-point contribution and therefore when carrying out calculations this term must be first subtracted to avoid double counting. The same unified model can be used to quantitatively describe the properties of ferroelectrics in both the quantum and classical regimes. Other quantities such as the heat capacity, thermal conductivity, soft-mode frequency and thermal expansion coefficient can also be calculated within the same framework. In certain limits which we now consider, closed-form analytical solutions are possible. Close to the quantum critical point where $\chi^{-1}$ is small and $k_B T/\hbar$ is less than the characteristic frequency, the inverse susceptibility is predicted to vary as temperature squared $\chi^{-1} = a + (10/3)(\varepsilon_0 k_B^2 b \Delta^2 / 12 \hbar a v^2) T^2$ and the corresponding soft-mode frequency $\Omega = \sqrt{(\Delta^2/a)\chi^{-1}}$ is found to vary linearly with temperature $\Omega_{q=0} \sim T$. At higher temperatures where $k_B T/\hbar$ is greater than the characteristic frequency, the large *T* approximation of the Bose function can be taken and the usual Curie-Weiss form of the inverse susceptibility is recovered $\chi^{-1} \sim T$ that is independent of any terms involving $\hbar$ as should be the case in the classical limit. In this region the soft-mode frequency varies as the square-root of temperature $\Omega_{q=0} \sim \sqrt{T}$. Finally for a paraelectric at low temperatures but further away from the quantum critical point, the gap in the dispersion relation is quite large $\Delta^2 \geq v^2 q^2$ and $k_B T/\hbar$ is still much less than the characteristic frequency. In this limit one can assume the soft-mode frequency is independent of wavevector and one finds
$\chi^{-1} = a + \lambda \left( (\hbar\Delta/2) coth(\hbar\Delta/2k_B T) - (\hbar\Delta/2) \right)$,
where $\lambda = 5 \varepsilon_0 b q_c^3 \hbar\Delta/18\pi^2 a$, which is constant at the lowest temperatures. This is the independent-ion or Einstein approximation and is also known as Barrett's formula[29]. It is sometimes used to make comparisons with measurements in quantum paraelectrics. We stress that this (non-self-consistent) equation is only applicable in a confined region of the phase diagram in the very low temperature limit, away from the quantum critical point on the paraelectric side. In other areas of the phase diagram and in particular close to the quantum critical point, it is crucial that the dispersion of the soft-phonon mode is taken into account. The schematic low temperature phase diagram for a cubic displacive ferroelectric is shown below in Fig. 1 and summarises the predictions of the model. Note that one does not expect an abrupt change in the temperature exponent $\gamma$ of the inverse susceptibility from a power law fit ($\delta\chi^{-1} \sim T^\gamma$) when moving between the different paraelectric regions but a gradual cross-over.

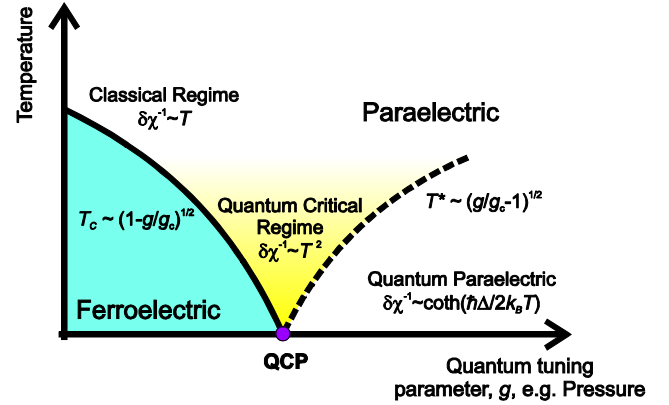

**Figure 1 | Schematic phase diagram of the ferroelectric quantum critical point.** The quantum tuning parameter used to suppress the transition towards zero temperature could be pressure or alternatively chemical or isotopic substitution. The blue ferroelectric region and paraelectric regions are separated by a second-order phase transition (solid line) terminating at a quantum critical point (QCP) at zero temperature. $\delta\chi^{-1}$ denotes the change in the inverse susceptibility from its zero temperature value. The dashed line and fading gradient shading indicate cross-over temperatures between the three paraelectric regimes that are described within the same unified model outlined in the text. The quoted temperature dependencies of the inverse dielectric susceptibility and cross-over temperatures are determined within a mean-field approximation and for the low temperature limit in the quantum paraelectric and quantum critical regime.

The preferred control parameter *g* is hydrostatic pressure since this can be used to cleanly vary the soft-mode interactions tuning $T_C$ to zero at a critical value $g_c$. Chemical or isotopic substitution on the other hand inevitably introduces disorder which may complicate the physics in the quantum regime. The model predicts that close to the quantum critical point, the Curie temperature varies with tuning parameter as $T_C \sim (1 - g/g_c)^{1/2}$ and the cross-over temperature $T^*$ between quantum critical paraelectric and conventional quantum paraelectric varies as $T^* \sim (g/g_c - 1)^{1/2}$.

Since we are at the marginal dimension ($d_{eff} = 4$) the mean field approximation used above is expected to be modified by logarithmic factors. This can be calculated in an extension of the model and has been determined in other works[15-17,19]. Close to the quantum critical point the $T^2$ form of the inverse susceptibility is adjusted to become $\chi^{-1} \sim T^2 (1 +$ $\mu ln(\rho/T))^{-1/3}$ where $\mu$ and $\rho$ are material specific constants and functions of the previously defined parameters of the model. Note that the logarithmic corrections cause only a very weak deviation from the $T^2$ dependence of the inverse susceptibility.

The model can also be extended to include the effects of long range dipole interactions in materials where the above assumptions are not appropriate such as the uniaxial case, damping of the phonon modes due to interactions with other phonons or itinerant electrons, coupling of the soft transverse-optic branch to other phonon branches or strain as well as to study quantum anti-ferroelectrics where an optical phonon becomes soft at a finite wavevector. Since order-disorder and displacive ferroelectrics can be understood within a unified microscopic model[9,30], the relevant generalization of the quantum model above can also be applied to order-disorder ferroelectrics such as $NaNO_2$ and intermediate ferroelectrics such as $KH_2PO_4$. In these cases damping of the modes as temperature increases should be considered.

**Results for $SrTiO_3$ and $KTaO_3$**

The incipient displacive ferroelectrics strontium titanate ($SrTiO_3$) and potassium tantalate ($KTaO_3$) with perovskite cubic crystal structures at room temperature are readily available as high-purity single crystals with lattice constants $l = 3.905$ Å and $l = 3.989$ Å respectively. They are chosen as materials in which to investigate the physics around the quantum critical point since they are naturally very close to a ferroelectric instability at absolute zero without the need to apply pressure or use other tuning apparatus and their conventional properties have been studied in detail in the past. The close proximity to the quantum critical point is evident from Raman[31], neutron[32,33] and dielectric experiments[14,34] where the susceptibility is observed to grow as much as $10^4$ at liquid helium temperatures despite remaining paraelectric down to the lowest temperatures measured. The materials are so close to an instability that small perturbations in the material or chemical environment are seen to induce ferroelectricity. For example, in $SrTiO_3$, elemental substitution[35] and strain[36] lead to finite temperature ferroelectricity. Fully substituting the oxygen-16 isotope with the oxygen-18 isotope results in a ferroelectric transition in $SrTiO_3$ at approximately 25 K[37]. The oxygen-18 substituted $SrTiO_3$ at ambient pressure sits to the left of the quantum critical point in Fig. 1. Since it is in an intermediate region of the phase diagram between the classical and quantum critical regime, one would not expect to observe either a $T^2$ or linear $T$ dependence of the inverse dielectric constant over an extended temperature range but perhaps an intermediate power law.

At approximately 105 K, $SrTiO_3$ undergoes a structural phase transition in which the oxygen octahedra rotate in opposite directions in neighbouring unit cells. A side effect of this is a small tetragonal distortion of the lattice that occurs at the same temperature. $KTaO_3$ on the other hand remains cubic down to low temperatures but is further from the quantum critical point having an estimated correlation length ξ, at 4 K of approximately 20 lattice constants compared to 40 lattice constants in $SrTiO_3$ at the same temperature.

Below we show for the first time a comparison of high precision measurements of the inverse dielectric constant as a function of temperature with the quantitative predictions of the self-consistent phonon model described above for real crystals close to a low temperature ferroelectric instability. By using the temperature independent parameters $a$, $b$, $\Delta$, $v$ quoted in Table 1 below, the model can be self-consistently solved at each temperature without any adjustable parameters. $a$ and $b$ were obtained by measuring the polarization in an electric field at 0.3 K and fitting the resulting plots with the zero temperature equation of state. The remaining parameters were obtained from inelastic neutron and Raman scattering experiments at 4 K.

|   | $a = \chi^{-1}$ | $b$ ($C^{-2}m^4$) | $\hbar\Delta/k_B$ (K) | $v$ ($ms^{-1}$) |
|---|---|---|---|---|
| $SrTiO_3$ | $5 \times 10^{-5}$ | 0.07 | 24 | 8090 |
| $KTaO_3$ | $2 \times 10^{-4}$ | 0.07 | 36 | 5690 |

**Table 1 | Material specific parameters for $SrTiO_3$ and $KTaO_3$.** $a$ an $b$ were obtained by measuring the polarization as a function of field up to 15 kVcm$^{-1}$ at 0.3 K and fitting to the zero temperature equation of state as *Ferroelectric Arrott plots*. The remaining parameters were determined from inelastic neutron[32,33] and Raman scattering[31] experiments at 4 K.

Note that precisely at the quantum critical point $g = g_c$, the gap $\Delta$ would be equal to zero. Both $SrTiO_3$ and $KTaO_3$ are close to the quantum critical point but on the paraelectric side and thus have a small but finite gap, $KTaO_3$ being slightly further away from the quantum critical point than $SrTiO_3$.

The comparison between the predictions of our model and experiment for the temperature dependent susceptibility of SrTiO$_3$ are presented in Fig. 2. We observe a good qualitative agreement between theory and experiment. Moreover, in terms of independently measured zero temperature parameters, the theory predicts the correct order of magnitude of the temperature dependent susceptibility. This agreement suggests that the model captures the essential physics of the paraelectric state close to the quantum critical point. Fig. 2a shows the experimental data. The high temperature inverse dielectric constant varies linearly with temperature with a Curie-Weiss constant of 8x10$^4$ K. Extrapolating the high temperature classical fit indicates an expected $T_C$ of 35 K, however at temperatures less than around 50 K, the inverse dielectric constant begins to flatten out. Below this cross-over temperature, the system enters a regime dominated by quantum critical fluctuations and using our high-precision data we highlight for the first time in the inset to Fig. 2a, the striking agreement to the simple model prediction, $\varepsilon^{-1} \sim T^2$. This regime is rather analogous to that in low temperature helium where classically one would expect to observe the solidification of helium but the existence of quantum fluctuations preserves the liquid state down to absolute zero.

The model result is shown in Fig. 2b over the same temperature range and closely predicts the quantum and classical regimes and correctly gives the temperature scale of the cross-over region. The quantum critical regime is predicted to be dominant when the temperature is much less than the Debye temperature of the critical phonon branch responsible for ferroelectricity, $\theta_{D,TO} = \hbar v q_c / k_B$. For SrTiO$_3$ $\theta_{D,TO} \approx 500$ K, and the cross-over to quantum behaviour is observed at about 50 K, i.e. 10% of $\theta_{D,TO}$. Fig. 2c shows the inverse susceptibility exponent γ against temperature from a power law fit ($\delta\chi^{-1} \sim T^\gamma$) of the experimental and model data in the same plot. The experimental curve in red shows a plateau region at γ = 2 (quantum critical regime) between about 8 K and 50 K before rising to a higher power at lower temperatures. This is because SrTiO$_3$ is not precisely at the critical value of the tuning parameter and the model predicts exponential type behaviour at the lowest temperatures in the 'conventional' quantum paraelectric regime (See Fig. 1). The model curve everywhere predicts a slightly higher power compared to experiment but is remarkably accurate given the simplicity of the model. The features marked with arrows on the experimental curve are not visible in the model. The peak labelled at 105 K is due to the antiferrodistortive phase transition in SrTiO$_3$ (arising from a soft phonon mode at the R point of the Brillouin zone). The origin of the feature at 10 K, which also shows up as an anomaly in the dielectric loss, is thought to be extrinsic, related to oxygen vacancies, and therefore not part of the present analysis (details given in ref. (38)).

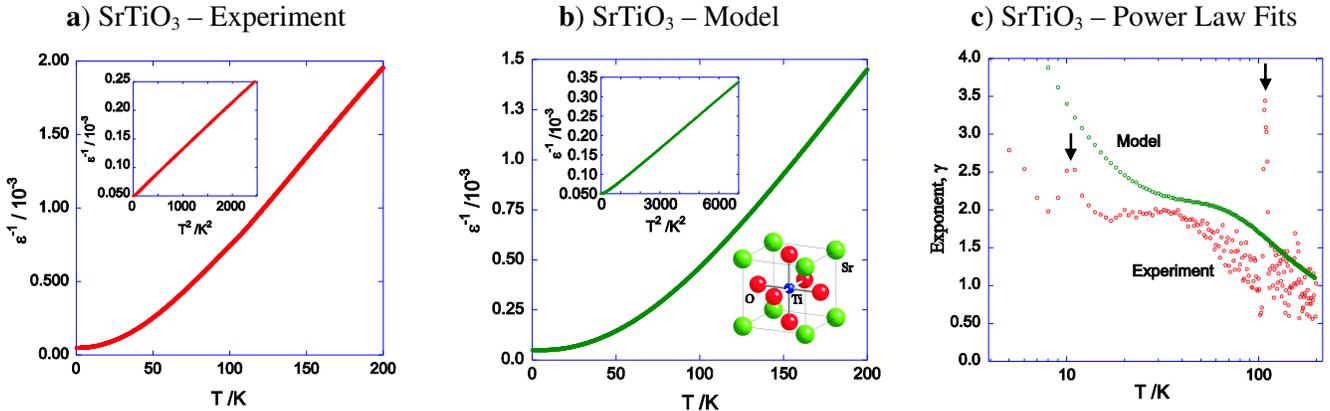

**Figure 2 | Experimental (a) and self-consistent phonon model results (b) for the inverse dielectric constant in SrTiO$_3$.** The upper left insets show the inverse dielectric constant plotted against temperature squared. The lower right inset in (**b**) shows the crystal structure of SrTiO$_3$. **c**, Figure showing the results of a power law fit ($\boldsymbol{\delta\chi^{-1} \sim T^\gamma}$) of the experimental and model data. The features marked with arrows are related to the structural transition in SrTiO$_3$ at 105 K and an effect at 10 K also observed in the dielectric loss tangent that is thought to be extrinsic and related to oxygen vacancies[38]. These features have not been incorporated into the model.

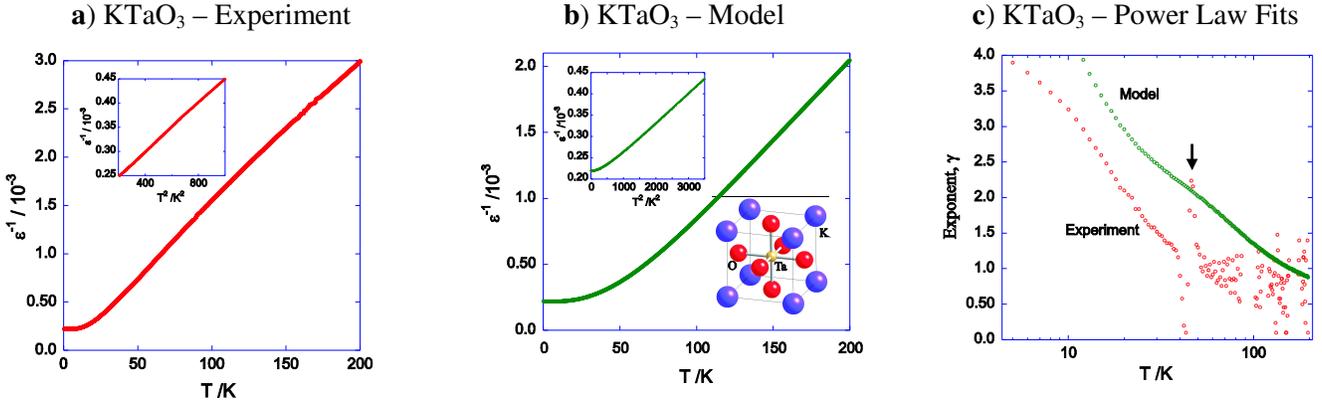

**Figure 3 | Experimental (a) and self-consistent phonon model results (b) for the inverse dielectric constant in KTaO$_3$.** The upper left insets show the inverse dielectric constant plotted against temperature squared. The lower right inset in (**b**) shows the crystal structure of KTaO$_3$. **c**, Figure showing the results of a power law fit ($\delta\chi^{-1} \sim T^{\gamma}$) of the experimental and model data. The feature marked with an arrow at 40 K is thought to be an extrinsic effect related to oxygen vacancies[38] and is also observed as a peak in the dielectric loss.

The results for KTaO$_3$ shown in Fig. 3 are broadly similar to those observed in SrTiO$_3$. The main differences arise because KTaO$_3$ is further from the quantum critical point and deeper in the paraelectric regime. Again a Curie-Weiss like linear dependence of $1/\varepsilon$ is seen at higher temperatures which flattens out with a higher power below around 30 K as quantum effects begin to dominate. Again the cross-over from classical to quantum behaviour is found at roughly 10% of $\theta_{D,TO}$ ($\theta_{D,TO} \approx 340$K). The key difference for KTaO$_3$ is that no extended region with $\gamma = 2$ is seen either in the model or experiment as shown in Fig. 3c. This is because the ground state of KTaO$_3$ is further from the quantum critical point than SrTiO$_3$ and sits more to the right of the QCP in the phase diagram shown in Fig. 1. Again the peak in the experimental curve in Fig. 3c at around 40 K, which also shows up as peak in the dielectric loss, is thought to be an extrinsic effect due to oxygen vacancies and not part of the present discussion.

The high precision data above provides an example of two materials, (1) SrTiO$_3$, which is in the quantum critical regime between around 8 K and 50 K and (2) KTaO$_3$, which is in the quantum paraelectric regime below around 30 K but does not show an extended region in temperature with quantum critical fluctuations. All the above results are well explained both qualitatively and quantitatively within the self-consistent phonon model summarised in Fig. 1. Even though the key physics is brought out within the above model, the differences between the model and experimental results, however small, could perhaps arise for the following reasons. Firstly the model only accounts for the contribution of a single phonon branch to the dielectric constant. The interaction between the polar transverse-optic phonons and other phonon branches have been left out. The order parameter expansion stops at the quartic level and no temperature dependence of the parameter *b* is assumed. Finally, the effect of long-range dipole interactions has not been quantitatively modelled but simply the longitudinal component of the fluctuating polarization has been removed. All of these effects are believed to be of secondary importance in the materials investigated here and are not required to understand the key physics.

The low temperature part of the measured inverse dielectric constant is magnified below in Fig. 4 where to our knowledge we reveal for the first time a broad minimum in both SrTiO$_3$ and KTaO$_3$ at 2 K and 3 K respectively (alternatively observed as a maximum in $\varepsilon$ against T). This is not seen in the model plots in Fig. 2b and Fig. 3b. These features keep the same shape, magnitude and position in temperature on varying the frequency over the range tested (1KHz to 1MHz). We reproduced the same results on a number of different samples and cryostats and the effect is most clearly seen using our high precision set-up. The percentage change in $\varepsilon$ from the position of the maximum to base temperature is larger in SrTiO$_3$ than in KTaO$_3$ and of the order 0.1%. Note that as is consistent with the third law of thermodynamics, the curves in Fig. 4 should enter the vertical axes at $T = 0$ K with zero slope. Indeed we found this to be the case when measuring the low temperature part of $1/\varepsilon$ down to 40 mK.

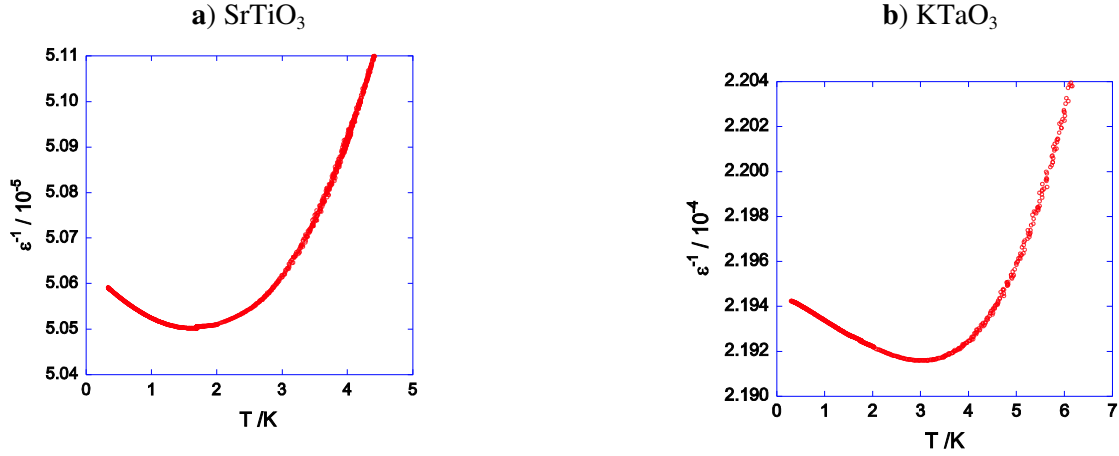

**Figure 4 | Observation of a minimum in the inverse dielectric constant in both a, SrTiO$_3$ and b, KTaO$_3$ in the neighbourhood of the quantum critical point on the paraelectric side.**

## Importance of Acoustic Phonons

It is known from earlier experiments in KTaO$_3$ and SrTiO$_3$ that the optical modes becoming critical at low temperatures interact with low lying transverse acoustic modes. This can be observed for example in inelastic neutron scattering data where the interaction can lead to the softening of the acoustic branch at finite wavevectors[39]. Coupling of the optical modes to acoustic modes can lead to a minimum in the inverse dielectric constant at low T when $T \ll \Delta$ (see Table 1 for values of $\Delta$). While the origin of the minimum shown in Fig. 4 is yet to be definitively determined, the interaction of the acoustic phonons (sound waves) with optical phonons is a promising candidate. This effect was first investigated theoretically by Khmelnitskii and Shneerson[16] in the early days of the study of quantum critical phenomena and has recently been reinvestigated in light of our new results by Pálová et al[20]. It involves the addition of a term $-\eta \nabla \varphi P^2$ in the free energy where $\varphi$ represents the (non-critical) acoustic phonons. The gradient of $\varphi$ is the strain and the electrostrictive coefficient $\eta$ controls the strength of the interaction. The correction in $\chi^{-1}$ due to this term is initially negative (varying as $-T^4$ in leading order in $T$), which produces the minimum. Similar physics is present in magnetic systems where strain couples to magnetisation with the same symmetry[40,41,42]. The crucial point is that to observe the emergence of a minimum such as the one we have found, a material must be close to the quantum critical point and on the paraelectric side at low temperatures. The relative positions and magnitudes of the minimum are consistent with the values of $\Delta$ and the speed of acoustic phonons for the two materials. Another consequence of coupling to long wavelength acoustic phonons is the possibility that the size of the parameter $b$ can be reduced and turned negative. This would mean that in certain regions of the phase diagram, for some materials, the phase transition line would have a pronounced first-order nature at low temperatures.

## Outlook

The close correspondence between theory and measurement of the dielectric constant presented in this paper suggests that the ferroelectrics can provide us with a particularly simple and perhaps text-book example of quantum criticality in the solid-state. The quantum critical regime arises from the lattice system alone and persists up to high temperatures of the order 50 K in the materials we have investigated. The simple self-consistent phonon model works well at explaining the key physics around the ferroelectric quantum critical point and also describes the cross-over to the classical regime. Moving away from the simplest cases brings a whole range of rich phenomena open for investigation. For example, looking for logarithmic corrections at the marginal dimension is feasible in these quantum critical systems. Coupling of the order parameter to sound waves (acoustic phonons) can lead to the emergence of a new low temperature regime in quantum ferroelectrics such as we have found in SrTiO$_3$ and KTaO$_3$ below liquid helium temperatures. In the case of multiferroics, e.g. EuTiO$_3$ [43] and others[44], coupling of the fluctuating polarization and magnetisation at low temperatures allows one to study coupled critical order parameters in the quantum regime. Also, another major avenue for research is to reintroduce conduction electrons into crystals within the ferroelectric quantum critical regime. Via the selection of particular materials, pressure tuning or doping of electrons, this can be done in a controllable fashion. Although the conduction electrons produce damping of the ferroelectric fluctuations and prevent the manifestation of an external electric field, below a certain screening length, conduction electrons would feel a local field due to the ferroelectric polarization[45]. Extra layers of complexity can be built into the system by design in order to study electron correlations mediated by the effective interactions at the quantum critical point. This is particularly promising in transition metal oxide interfaces[46,47] where metallic layers are created between for example two insulating perovskite lattices. Alternatively, doping of electrons into the bulk can be carried out in $SrTiO_3$ which becomes superconducting at low temperatures[48]. Other ferroelectrics are naturally conducting such as the low conductivity metals $SnTe$[49] and $GeTe$[50] that both exhibit superconductivity at low temperatures. All of these promising areas of research and others will be greatly aided by the concepts of quantum criticality in ferroelectrics as discussed here.

**Methods**

A number of samples of $SrTiO_3$ and $KTaO_3$ were used in these experiments. All were single crystals and cut into rectangular slabs of approximate dimensions 3mm × 3mm × 0.5 mm orientated in the [100] direction. Dielectric measurements were carried out in the parallel plate capacitance geometry by evaporating gold electrodes on the large surfaces. Two cryostats, an adiabatic demagnetisation refrigerator (40 mK to 300 K) and a pumped helium-3 system (0.3 K to 300 K) both fitted with miniature coaxial cables were used for high precision signal detection with an Andeen Hagerling and QuadTech capacitance bridges. Measurements were taken during heating at a rate of less than 1 K per hour at low temperatures and approximately 5 K per hour at high temperatures. The measurement frequency could be varied between a few Hz and a few MHz. It was possible to apply low noise DC fields up to 20 kVcm$^{-1}$ during measurements.

1. Laughlin, R. B. and Pines, D. The theory of everything. Proc. Natl. Acad. Sci. USA 97, 28–31 (2000).
2. Scott J.F. Science 315, 954-959 (2007)
3. Smith, R. P. et al, Nature 455, 1220-1223 (2008)
4. Pfleiderer, C. et al, Nature 427, 227-231 (2004)
5. Lonzarich, G.G., Montoux, P. and Pines, D., Nature 450, 1117, (2008)
6. Mathur, N.D. et al, Nature 394, 39 (1998)
7. Saxena, S.S. et al, Nature, 406, 587, (2000)
8. Ahlers, G. et al, Phys. Rev. Lett, 34 No 19 (1975)
9. Strukov, B.A. and Levanyuk, A., *Ferroelectric Phenomena in Crystals*, Springer-Verlag (1998)
10. Lines, M.E. and Glass, A.M., *Principles and Applications of Ferroelectrics*, Oxford University Press (2004)
11. Cowley, R. A., Adv. Phys. 29, 1 (1980); Bruce, A. D., Adv. Phys. 29, 111 (1980); Bruce, A. D. and Cowley, R. A., Adv. Phys. 29, 219 (1980).
12. King-Smith, R.D.and Vanderbilt, D., Phys. Rev. B 47, 1651, (1993)
13. Resta, R., Rev. Mod. Phys. 66, 899 (1994)
14. Müller, K.A. & Burkard, H., Phys. Rev. B,19, No. 7 (1979)
15. Rechester, A.B., Sov. Phys, JETP 33, 423, (1971)
16. Khmelnitskii, D. E. and Shneerson, V. L., Sov. Phys. – Solid State, 13, 687 (1971)
17. Khmelnitskii, D. E. and Shneerson, V. L., JETP 37, 164, (1973)
18. Schneider, T., Beck, H. and Stoll, E., Phys. Rev. B, 13, 1123, (1976)
19. Roussev, R. and Millis, A.J., Phys. Rev. B 67, 014105 (2003)
20. Pálová, L., Chandra, P., and Coleman, P., Phys. Rev. B 79, 075101 (2009)
21. Lonzarich, G.G. and Taillefer, L., J. Phys. C: 18, 4339 (1985)
22. Lonzarich, G.G., Chapter *The Magnetic Electron*, in *Electron* Edited Springford, M., Cambridge University Press (1997)
23. Moriya, T., *Spin Fluctuations in Itinerant Electron Magnetism*, Springer, Berlin, (1985)
24. Sachdev, S, *Quantum Phase Transitions*, Cambridge University Press (2001)
25. Larkin, A.J. and Khmelnitskii, D.E., JETP 29, 1123, (1969)
26. Sandvold, E. and Courtens, E., Phys. Rev. B 27, 5660, (1983)
27. Martinez, J. and Gonzalo, J., Phys. Rev. B 32, 400, (1985)
28. Landau, L.D. and Lifshitz, E.M., *Statistical Physics Part 1*, Butterworth Heinemann (2002)
29. Barrett, J. H., Phys. Rev. 86, 118 (1952)
30. Rubtsov, A.N. and Janssen, T., Phys. Rev. B 63, 172101 (2001)
31. Fleury, P.A. and Worlock, J.M., Phys. Rev. 174, 613 (1968)
32. Yamada, Y. and Shirane, G., J. Phys. Soc. Japan 26 396 (1969)
33. Shirane, G., Nathans, R. and Minkiewicz, V.J., Phys. Rev. 157 396-399 (1967)
34 Wemple, S. H., Phys. Rev. 137, A1575 (1965)
35. Bednorz, J. G. and Müller, K. A., Phys. Rev. Lett. 52, 2289 (1984)
36. Haeni, J.H. et al, Nature 430, 758 (2004)
37. Itoh, M.et al, Phys. Rev. Lett. 82, 3540 (1999)
38. Scott, J.F., Journal of Physics-Condensed Matter 11, 8149 (1999)
39. Axe, J.D. et al, Phys. Rev. B 1, 1227 (1970)
40 Rice, O.K., J. Chem. Phys. 22, 1535 (1954)
41 Larkin, A.I. and Pikin, S.A., JETP 29, 891 (1969)
42. Green, A .G. et al, Phys. Rev. Lett. 95, 086402 (2005)
43 Spalek, L.J., Shimuta, M., Rowley, S.E., Katsufuji, T., Petrenko, O. A., Saxena, S.S. and Panagopoulos, C., Proceedings of SCES'08
44 Kim, J.W. et al, arXiv:0810.1907v2 [cond-mat.mtrl-sci]
45. Anderson, P.W. and Blount, E.I., Phys. Rev. Lett. 14, 217 (1965)
46. Okamoto, S. and Millis, A. J., Nature 428, 630 (2004)
47 Ohtomo, A. and Hwang, H. Y., Nature 427, 423 (2004)
48. Schooley, J.F. et al, Phys. Rev. Lett. 14, 305 (1965)
49. Iizumi, M. et al, J. Phys. Soc. Jap. 38, 443, (1975); Kobayashi, K.L.I. et al, Phys. Rev. Lett. 37, 772, (1976)
50. Baghat, A.A. et al, Physica B 382, 271 (2006); Chatopadhyay, T. et al, J. Phys. C: Solid State Phys. 20, 1431 (1987)

**Acknowledgments**

We would like to thank D.E. Khmelnitskii, S.J. Chorley, G. Catalan, I.R. Walker, P.B. Littlewood, C. Panagopoulos, L. Pálová, P. Chandra, P. Coleman, N. Marcano, N.D. Mathur and K.H. Kim for their help and discussions.

We would also like to acknowledge support from Emmanuel College Cambridge, the Engineering and Physical Sciences Research Council (EPSRC) of the United Kingdom and the European Research Council (ESF) COST - P16 programme.

**Author Information**

Correspondence and requests for materials should be addressed to S.E.R. (ser41@cam.ac.uk) and S.S.S. (sss21@cam.ac.uk).